\documentclass[aps,twocolumn,prl,color,psfig,epsf,showpacs]{revtex4}
\usepackage{amsmath}
\usepackage{amsfonts}
\usepackage{hyperref}
\usepackage{graphicx}
\begin{document} 
\title{Role of fluctuations in membrane models: thermal versus non-thermal}
\author{Amit K Chattopadhyay}
\address{Aston University, Non-linearity and Complexity Research Group, EAS, Birmingham B4 7ET, UK}
\email{A.K.Chattopadhyay@aston.ac.uk} 
% 
%\date{} 
% 
\begin{abstract} 
\noindent 
We study the comparative importance of thermal to non-thermal fluctuations for membrane-based
models in the linear regime. Our results, both in 1+1 and 2+1 dimensions,  
suggest that non-thermal fluctuations dominate thermal ones only when the 
relaxation time $\tau$ is large. For moderate to small values of $\tau$, the dynamics
is defined by a competition between these two forces. The results
are expected to act as a quantitative benchmark for biological modeling in systems
involving cytoskeletal and other non-thermal fluctuations.
\end{abstract} 
\pacs{05.70.Ln, 05.40.-a, 87.16.-b} 

\maketitle 

\section{Introduction}

\par
A long standing tradition in the field of non-equilibrium statistical mechanics, especially those 
in connection with biophysical dynamics \cite{background_references1, background_references2,background_references3,background_references4} has been to look 
at the high temperature limit of all such models where thermal fluctuations dominate every other
stochastic mode in the system. Such an approach generally has the advantage of
having to deal only with a thermal noise that dominates all non-thermal motions in the 
mesoscopic limit. Both theoretically 
\cite{background_references1,background_references2,background_references3,background_references4} and experimentally \cite{expt_thermal1, expt_thermal2,expt_thermal3,expt_thermal4}, 
the subject is replete with such examples, and often with
agreeable results too, where such an assumption agrees with the physiology of the real system.  
Problems, though, surface when the approach contradicts its basic tenet in that the system
has a non-trivial non-thermal force that competes, and sometimes even out-competes, the
thermal fluctuations at the experimental setting \cite{expt1, expt2, expt3, expt4}. In the
present article, our focus is on the latter situation when the system has a non-trivial 
component of non-thermal fluctuation eg. cytoskeletal fluctuation \cite{expt1,expt2,expt3} 
whose contribution can not be neglected in comparison to its thermal counterpart.

\par
A similar, if not identical, question in connection is whether or not there exists a non-equilibrium version of the fluctuation-dissipation (FD) theorem in the mould of an Jarzynski relation \cite{jarzynski1,jarzynski2,jarzynski3}. Our approach, though, will rely on a relative comparison of strengths between thermal and non-thermal (stochastic) fluctuations connected to identical reservoirs. \\

\par
A remarkable range of such non-thermal fluctuations are visible in the biological world, a typical case in hand being the fluctuations due to bio-chemical and molecular movements, popularly referred to as 'cytoskeletal fluctuations'. Recent studies \cite{cyto} have traced the origin of such (generally) non-thermal cytoskeletal dynamics to a combination of cell crawling, adhesion between lipid bilayers and randomness in the responses of molecular motor proteins. In this article, though, we would not pay so much of attention as to the origin of such motions; rather our focus would be on a relative comparison of strengths between a Brownian thermal noise against a randomness that is mostly non-thermal in origin eg. cytoskeletal fluctuations and whose strength is proportional to $\sqrt{T_{\mathrm{eff}}}$, $T_{\mathrm{eff}}$ being a 'non-equilibrium temperature'. The Langevin and Fokker-Planck type models based
on such a description \cite{background_references2, background_references3, akc_burroughs} 
generally utilize the fact that the amplitude of thermal motion thoroughly
dominates all other mesoscopic level fluctuations that are otherwise present in the dynamics.
Although such an approach works remarkably well in many cases
\cite{expt_thermal1,expt_thermal2}, and we are mostly focusing on native biological systems
involving cytoskeletons, serious complications arise in cases where the cytoskeletal
fluctuations are non-negligible. The simple question one would like to ask here is the
following: what are those situations when such a non-thermal motion competes with a thermal
fluctuation attached to a heat bath \cite{Lubensky}? A
notable contribution to the understanding of such non-thermally initiated fluctuations
owes to Prost and Bruinsma \cite{Prost_Bruinsma}. They showed that in systems involving
active membranes the long-wavelength limit of the fluctuation spectrum follows a power law
behavior that is quite different from a membrane in thermal equilibrium. Using this
approach they were successful in analyzing the experimental results on red-blood cells 
\cite{expt4} where cytoskeletal fluctuations were convincingly seen to be dominating
the thermal ones. Similar studies dealing with the effect of non-thermal motions 
in the intra-cellular trafficking involving membrane fluctuations \cite{Sriram2001}, or in the
travelling waves observed due to protein activity in a flexible membrane \cite{Prost2000},
as well as in micropipet experiments using an activated membrane surface \cite{Prost2001}
lend credence to the belief that active non-thermal motion can indeed have a dramatic
effect on the overall dynamics of a non-equilibrium system, especially when there is a 
competition with a fluctuations arising from a stochastic thermal bath.

\par
In this article, we study the regime when the two noise modes,
thermal and non-thermal, compete with each other. Ramifications of such comptetitions are amply evident in experiments \cite{expt1,expt2,expt3,expt4,Prost2001,Hac2005}
and our attention here is to quantify the limits of the parameter that defines such a regime.
As we would later see, the parameter in question is the relaxation
time $\tau$ that, in a way, offers a complementary description to that studied in 
\cite{Prost_Bruinsma}. 
We study the model in both asymptotic limits of $\tau$ and predict 
values (indeed limits) of $\tau$ for which thermal noise dominates (or competes with) 
its non-thermal counterpart.

\noindent
Our basic model consists of a large flat d+1 dimensional membrane driven by thermal 
fluctuations where the free energy is Helfrich-like \cite{Helfrich}, involving a term that is proportional to the energy cost due to surface tension and another term that represents the bending energy associated with the existence of curvature in a membrane. In our nomenclature, we will represent the surface tension term as $F_{\mathrm{surf}}=\frac{T}{2} \int d^dr {({\vec \nabla} Z)}^2$ and a curvature term
$F_{\mathrm{bend}}=\frac{B}{2} \int d^dr {({\vec \nabla}^2 Z)}^2$. However, in the present article, we 
focus only on the linear regime of such an intrinsically non-linear potential (in the mould
of \cite{akc_burroughs}). As can be easily seen, this linear regime can be arrived at as a 
high-temperature expansion of the Kramer's type two-state
potential studied in \cite{background_references1, background_references2}. As is always true of linear models, ours too would be constrained by the fact that exact measurements of amplitudes of perturbations as well as enacting proper boundary conditions would suffer. However, the gross aim being a qualitative understanding of the role of noise (both thermal and non-thermal) in perturbed membranes, we forego this without any effective loss of generality. \\

\par
\noindent
Our interest is in the activated dynamics of this membrane in the presence of
a thermal, together with a non-thermally fluctuating noise. We define the thermal part of the
noise using the standard FD theorem, while the non-thermal part includes a time-decaying component. 
The finite decay in the non-thermal noise distribution simply indicates the
presence of a finite relaxation time $\tau$ as opposed to a an infinite relaxation time
for a white noise source. One might consider more complicated noise sources as well, like
a non-local spatial correlation in its distribution. We, however, stick to the minimalist
model since a spatial non-locality does not change the qualitative outcome, apart from 
complicating the algebra that is. In the following description, the thermal noise
is represented by $\eta_{\text{th}}(\vec x,t)$ and the non-thermal noise by $\eta_{\text{nth}}(\vec x,t)$.
The over-damped dynamics of such a membrane
(we consider both 1+1 and 2+1 dimensional systems individually to illustrate the
generality of the argument)
defined by a stiffness $B$, surface tension $T$ and mobility $\gamma$ is given by

\begin{eqnarray}
\gamma \frac{\partial Z(\vec x,t)}{\partial t} &=& -B{\nabla}^4 Z(\vec x,t) + T {\nabla}^2 
Z(\vec x,t) \\ \nonumber 
&+& \mathrm{nonlinear\:\:terms} + F_{\mathrm{ext}}
\end{eqnarray}

\noindent
$F_{\mathrm{ext}}$ is the space-time dependent external force impressed on this system that,
in principle, can be stochastic. Defined generally, we define the force as a monotonically 
decaying function of time as follows

\begin{equation}
< F_{\mathrm{ext}}(\vec x,t) F_{\mathrm{ext}}(\vec x',t') > = F_0 \exp[-\frac{|t-t'|}{\tau}]
\delta(\vec x-\vec x')
\label{force_correlation}
\end{equation}

\par
\noindent 
Our objective is to study the spatio-temporal properties of this model 
both in 1+1 and 2+1 dimensions
for the two asymptotic regimes $\tau \rightarrow 0$ and $\tau \rightarrow \infty$
and compare such responses with a stochastically (thermal) driven string/membrane. 

\section{Non-thermal fluctuations in a 1+1 dimensional string}

\noindent
We start from a linearized version of a (1+1 dimensional) membrane model driven by a combination of an external time dependent {\it non-thermal} noise $\eta_{\text{nth}}$ and a Gaussian thermal noise $\eta_{\text{th}}$. 

\begin{equation}
\gamma \frac{\partial Z(\vec x,t)}{\partial t} = -B{\nabla}^4 Z(\vec x,t) + T {\nabla}^2
Z(\vec x,t) + \eta_{\text{nth}} +  \eta_{\text{th}}
\label{combined}
\end{equation}

\noindent
where the noises $\eta_{\text{nth}}$ and $\eta_{\text{th}}$ are defined as follows:

\begin{eqnarray}
<\eta_{\text{nth}}(\vec x,t) \eta_{\text{nth}}(\vec x',t')> &=&  F_0\:e^{-\frac{|t-t'|}{\tau}}\:\delta(\vec x - \vec x') \\ \nonumber
<\eta_{\text{th}}(\vec x,t) \eta_{\text{th}}(\vec x',t')> &=&  D_0\:\delta(t-t')\:\delta(x-x') \\ \nonumber
<\eta_{\text{nth}}(\vec x,t) \eta_{\text{th}}(\vec x',t')> &=& 0 \label{noises}
\end{eqnarray}

\noindent
Since the thermal and non-thermal noises are uncoupled \cite{uncoupled} (although not mutually exclusive; in fact the main aim of our study is to ascertain the dominant mode when both are simultaneously present), we will decouple equation (\ref{combined}) in to separate thermal and non-thermal modes without any loss of generality. Note that this is possible only when the non-thermal amplitude is completely independent of ('non-equilibrium') temperature as is the case in most biological systems. For exact quantifiation of the numbers to be evaluated henceforth, we would use parameter values as in \cite{akc_burroughs}: B=11.8 $k_B T$, T=5650 $k_BT$ $\mu m^{-2}$, $\gamma=4.7 \times {10}^6\:k_BT\:s\:\mu m^{-4}$ and $D_0=2k_BT\:M$. \\

\noindent
Our working model for the {\it non-thermal case} would then be

\begin{equation}
\gamma \frac{\partial Z(\vec x,t)}{\partial t} = -B{\nabla}^4 Z(\vec x,t) + T {\nabla}^2
Z(\vec x,t) + \eta_{\text{nth}}
\label{nonthermal}
\end{equation}

\noindent
where $F_0 \sim 1/\tau$ due to normalisation constraint. 
Solving the above equation in the wave vector space, we evaluate the two-point temporal
correlation function as 

\begin{eqnarray}
&<& {\tilde Z}_{\vec k}(t) {\tilde Z}_{-\vec k}(t') > = 
\frac{F_0}{\gamma^2 [\alpha(\vec k)-\frac{1}{\tau}]} \\ \nonumber
&\times&  \{ \frac{\exp[-(\frac{t-t'}{\tau})]}{\alpha(\vec k)+\frac{1}{\tau}} 
- \frac{1}{2\alpha(\vec k)} \exp[-\alpha(\vec k)(t'-t)] \}
\label{kspace_nonthermal}
\end{eqnarray}

\noindent
for $t'>t$, where $\alpha(\vec k)=\frac{B k^4+T k^2}{\gamma}$. This gives us the detailed
two-point structure function in 1+1 dimensions as

\begin{eqnarray}
&<& Z(\vec x,t) Z(\vec x',t') > =\frac{\gamma^{-5/4} \tau^{-1/4} B^{1/4}} 
{4T} \\ \nonumber 
&\times& \exp[-\frac{(t'-t)}{\tau}] 
(\frac{\exp[-X\sqrt{-\mu_{+}}]}{\sqrt{-\mu_{+}}}-
\frac{\exp[-X\sqrt{-\mu_{-}}]}{\sqrt{-\mu_{-}}}) \\ \nonumber
&+& \frac{1}{2\gamma^2} \int_{k_0}^{\infty}\:\frac{dk}{2\pi}\:\frac{e^{ikX-\alpha(\vec k)|t'-t|}}{\alpha(\vec k)}
\label{correlation_function}
\end{eqnarray}

\noindent
where $\mu_{\pm}=\frac{1}{2}[-\frac{T}{\sqrt{\epsilon B}} \pm \sqrt{\frac{T^2}{\epsilon B}-4}]$,
with $\epsilon=\gamma/\tau$ and $X={(\frac{\gamma}{B \tau})}^{1/4}|x-x'|$. $k_0$ relates to the smallest length scale in the system, which, in effect, turns out to be the lattice size $\lambda$ ($k_0=\frac{2\pi}{\lambda}$). \\

\noindent
The above equation specifies the two-point temporal auto-correlation function defined at the same spatial point (X=0) as

\begin{eqnarray}
&<& Z_{\text{nth}}(\vec x,t) Z_{\text{nth}}(\vec x,t') > = \frac{\gamma^{-5/4} \tau^{-1/4} B^{1/4}}
{4T} \\ \nonumber
&\times& (\frac{1}{\sqrt{-\mu_{+}}}-\frac{1}{\sqrt{-\mu_{-}}}) \exp[-\frac{(t'-t)}{\tau}] \\ \nonumber
&+& \frac{1}{2\gamma^2} \int_{k_0}^{\infty}\:\frac{dk}{2\pi}\:\frac{e^{-\alpha(\vec k)|t'-t|}}{\alpha(\vec k)}
\label{temp_correlation_function}
\end{eqnarray}

\noindent
where in the above equation we have replaced $Z$ by $Z_{\text{nth}}$ to indicate that it is the auto-correlation function for the non-thermal scenario.\\

\noindent
We now move on to a similar
looking model as defined in eqn (4) but with a thermal noise 
$\eta_{\text{th}}$ now replacing the non-thermal noise $\eta_{\text{nth}}$:

\begin{equation}
\gamma \frac{\partial {\hat Z_{\text{th}}}(\vec x,t)}{\partial t} = -B {\nabla}^4 {\hat Z_{\text{th}}}(\vec x,t) + T{\nabla}^2 {\hat Z_{\text{th}}}(\vec x,t) + \eta_{\text{th}}(\vec x,t)
\label{stochastic_equation}
\end{equation}

\noindent
where $\eta_{\text{th}}$ is as defined in eqn (\ref{noises}).
\noindent
Proceeding as before, the two point structure function for the {\it thermally perturbed} membrane in the Fourier transformed k-space reads as

\begin{eqnarray}
&<& {{\tilde Z}}_{\text{th}}(\vec k,t) {{\tilde Z}}_{\text{th}}(-\vec k,t') > = 
\frac{D_0}{\gamma^2}\:\frac{e^{-\alpha(\vec k)|t'-t|}}{2\alpha(\vec k)}
\label{kspace_thermal}
\end{eqnarray}

\noindent
In the limit $X \rightarrow 0$ and $|t'-t| \rightarrow \infty$, 
that is for very close spatial points evolved until the system reaches the stationarity limit, we get

\begin{equation}
 <{\hat Z_{\text{th}}}(\vec x,t) {\hat Z_{\text{th}}}(\vec x,t') >
= \frac{D_0}{\gamma^2}\:\:\int_{k_0}^{\infty}
\frac{dk}{2\pi} \frac{\exp[- \alpha(\vec k)(t'-t)]}{\alpha(\vec k)}
\label{thermal_eq}
\end{equation}

\noindent
Equation (\ref{thermal_eq}) above resembles the second part of equation (8) (eqn (\ref{thermal_eq}) approaches eqn (8) in the limit $\tau \to 0$) which implies that it is essentially the first part of the same equation (8) that will decide which noise dominates the system when both are simultaneously present. We would revisit this question soon after solving eqn (\ref{thermal_eq}) in the stationary state limit ($|t'-t| \to \infty$). \\ 

\noindent
Let us define $J=\int_{0}^{\infty}
\frac{dk}{2\pi} \frac{\exp[\alpha(\vec k)(t'-t)]}{\alpha(\vec k)}$. This can be represented as the following integral equation

\begin{equation}
\frac{\partial J}{\partial |t'-t|} = \int \frac{dk}{2\pi} e^{-\alpha(\vec k)|t'-t|}
\end{equation}

\noindent
which can be approximately solved in the limit $k_0 \to 0$ to get

\begin{eqnarray}
\frac{\partial J(|t'-t|)}{\partial |t'-t|} &=& 3\frac{e^{-\pi}}{4} \sqrt{\frac{\pi}{2}} \sqrt{\frac{T}{B}} e^{\frac{T^2}{8B\gamma}(t'-t)} \\ \nonumber
&\times&  K(1/4,\frac{T^2}{8B\gamma}(t'-t)) 
\end{eqnarray}

\noindent
where $K(n,x)$ represents the $n^{\text{th}}$ order Bessel function of the second kind. In the $|t'-t| \to $ large limit, the above expression reduces to

\begin{equation}
J(|t'-t|) = 3\frac{e^{-\pi}}{8} \sqrt{\frac{\gamma}{B|t'-t|}} \text{Erf}(\frac{T}{\sqrt{8 B \gamma}} \sqrt{|t'-t|}) 
\label{stochastic_temporal_correlation_function}
\end{equation}
 
\noindent
Erf(x) represents the incomplete Gamma function (also known as the error function) that is represented as $\text{Erf}(x)=\frac{2}{\sqrt{\pi}}\int_0^{x}\:du\:e^{-u^2}$. 

\begin{figure}
\centering
\includegraphics[scale=0.3,angle=0]{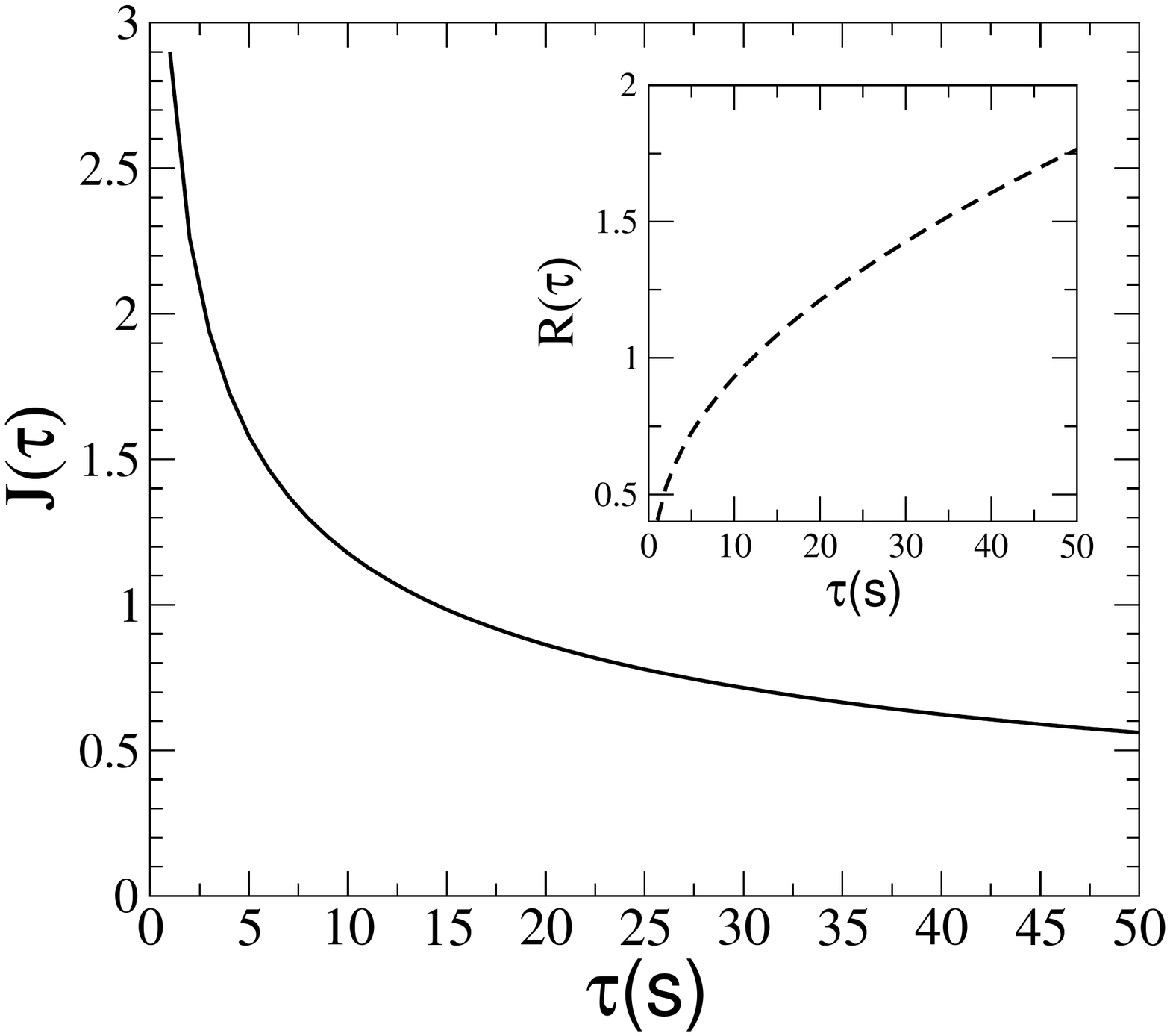}
\caption{The outset shows the variation of $J(|t'-t|)$ with varying relaxation time $\tau$ for $|t'-t| \sim \tau$ while the inset shows the ratio of the non-thermal to thermal contribution $R(\tau)$.}
\label{fig_Rplot}
\end{figure}

\noindent
The comparative strength of the non-thermal versus the thermal noise is encapsulated in the ratio $R=\frac{<Z_{\text{nth}}(x,t) Z_{\text{nth}}(x,t')>}{\frac{D_0}{\gamma^2}J(|t'-t|)}$ as could be evaluated from equations (8) and (13).  With the parameter values previously specified, $\frac{1}{2 \gamma^2}{J(|t'-t|)}|_{|t'-t| \to \tau} \sim {10}^{-12}$, implying $R>0$ (inset of the figure). As can be clearly seen from Fig. {\ref{fig_Rplot}} (reminiscent of experimental observations too \cite{Hac2005}), in the limit of large $\tau$, J becomes independent of the relaxation time while R too starts to saturate. At this point ($\tau \to$ large), non-thermal noise starts taking over the thermal fluctuations. With the parameter values used, $\tau$ is of the order of a few seconds (as shown in the figure), although much larger (~hours) or smaller (~ns) relaxation times are not unknown \cite{Hac2005}. In the limit of smaller relaxation times, that is transient perturbations, the non-thermal (eg. cytoskeletal) fluctuations compete against all thermal modes. \\

\section{Non-thermal fluctuations in a 2+1 dimensional membrane}

\noindent
Taking cues from the linearized model for non-thermal fluctuations eqn (\ref{nonthermal})
and its eventual k-space equivalent, we arrive at 

\begin{eqnarray}
&<& {\tilde Z}_{\vec k}(t) {\tilde Z}_{-\vec k}(t') > =
\frac{F_0}{\gamma^2 [\alpha(\vec k)-\frac{1}{\tau}]} \\ \nonumber
&\times&  \{ \frac{\exp[-(\frac{t-t'}{\tau})]}{\alpha(\vec k)+\frac{1}{\tau}}
- \frac{1}{2\alpha(\vec k)} \exp[-\alpha(\vec k)(t'-t)] \}
\label{kspace_nonthermal}
\end{eqnarray}

\noindent
As in the previous section, we utilize this information to evaluate the 2+1 dimensional
structure function

\begin{eqnarray}
&<& Z(\vec x,t) Z(\vec x',t') > \\ \nonumber 
&=& \int \frac{d^2k}{{(2\pi)}^2}\:e^{ikX \cos{\theta}}\:<{\hat Z}_{\vec k}(t) {\hat Z}_{-\vec k}(t)> \\ \nonumber
&=& \frac{1}{4\pi \tau \gamma^2} \int\:dk\:k\:J_0(kX)
\{ \frac{e^{-\frac{|t'-t|}{\tau}}}{\alpha(\vec k)+\frac{1}{\tau}} \\ \nonumber 
&+& \frac{1}{2\alpha(\vec k)}e^{-\alpha(\vec k)(t'-t)} \}
\label{membrane_structure_function}
\end{eqnarray}

\noindent
where $J_0(kX)$ is the zeroth order Bessel function of the first kind. Resorting to the
usual limit $X \rightarrow 0$, we get $J_0(kX) \rightarrow 1 - \frac{k^2 X^2}{4} +  
O(k^4 X^4)$. As can be easily seen, eqn (\ref{membrane_structure_function}) is perfectly
integrable for the first term in this equation while the second term shows a mild logarithmic
divergence that can be taken care of by assuming a minimum cut-off length scale 
for the system, a length that simply represents the lattice size of the discretized system. 
The basic conclusion remains
unchanged though, that for large values of the relaxation time $\tau$ the non-thermal fluctuations
will dominate the thermal ones while for small values of $\tau$ the 
dynamics will be decided by a competition between these two. \\

\section{Conclusions}
In this paper, we have discussed the relative importance of thermal noise with respect to non-thermal
fluctuations. The results give us a clear quantitative basis for considering or neglecting non-thermal
fluctuations in representative biological models. This is of utmost importance, since a traditional
disturbing tendency in such studies has often been to neglect non-thermal modes in favor of thermal 
noise. Although, this entails the disconcerting possibility of having to consider contributions 
from an often less understood source of fluctuation, like cytoskeletal fluctuations in the analyzes 
of microtubule related studies, an advantage on the hindsight is that of a theoretical clarity.
A knowledge of the relaxation mechanism that hopefully would be available from the experimental studies
would be a huge boost to the complementary theoretical modeling of such systems. Work is presently underway to  evaluate quantitative modifications in the presence of competing non-linearities in the dynamics. \\

\section{Acknowledgments}
AKC acknowledges partial research support from the Marie Curie Incoming
International Fellowship MIFI-CT-2005-008608 during the initial phase of this work. Matthew Turner is also acknowledged for inciteful discussions during the course of this work.

\end{document}